# The development of vibration modes propagation method to perform wave-optics simulation of beamline vibration

Han Xu[1], Xiao Li[1,2], Ming Li[1], Zhe Ren[1], Yi Zhang[1], Peng Liu[1], Yuhui Dong,[1,3] and Liang Zhou[1,4]

[1]*Beijing Synchrotron Radiation Facility, Institute of High Energy Physics, Chinese Academy of Sciences, Beijing, People's Republic of China.*

[2]*School of Nuclear Science and Technology, University of Chinese Academy of Sciences, Beijing 100049, China*

[3]dongyh@ihep.ac.cn

[4]zhouliang@ihep.ac.cn.

**Abstract**

The evolution from 3$^{rd}$ to 4$^{th}$ generation synchrotron radiation (SR) sources provide promising potential improvements in X-ray techniques, particularly in spatial resolution for imaging, temporal resolution for dynamic studies, and beam size control for nanoprobes. Achieving these enhancements demands effective vibration suppression in beamline systems. This challenge drives the need for optical designs that ensure efficient photon transport while maintaining vibration within acceptable thresholds. To address the advanced coherence requirements of fourth-generation SR sources, wave-optics simulations must be incorporated into optical design processes. We therefore propose a vibration mode propagation method using wave-optics techniques for beamline vibration simulation. Our approach achieves an almost 40-fold computational acceleration in actual beamline models compared to conventional methods, enabling direct analysis of propagating wavefront vibrations. This framework allows systematic evaluation of intensity distribution variations, coherence changes, and beam positioning errors caused by mechanical vibrations.

# 1. Introduction

The transition from 3rd to 4th generation synchrotron radiation (SR) sources demands high vibration control in beamline systems, driven by the ultra-low emittance of the source [1] and the emerging advanced X-ray techniques: (1) vibration amplitudes usually required to match the spatial resolution targets in X-ray imaging [2, 3], (2) temporal stability must exceed experimental durations in dynamic studies like X-ray photon correlation spectroscopy (XPCS) experiment [4, 5], (3) the decoherence effect [6, 7, 8] of the high frequency mechanical vibrations could prohibit the application of coherent scattering techniques such as coherent diffraction imaging (CDI), and (4) X-ray nanoprobe applications utilizing advanced optics such as multilayer Laue lenses (MLLs) and phase-correction plates, demand vibration amplitudes below 1 nm to maintain sub-10 nm beam sizes (10% ). Such technical demands necessitate both optimized beamline designs and advanced simulation methods incorporating wave-optics principles.

Various theoretical and computational approaches have been developed for beamline vibration analysis. Goto et al. [9] established a phase-space model that characterizes optical vibrations through statistical phase-space distributions, providing valuable insights into vibration-induced beam parameter variations. Shi et al. [7] proposed a virtual source model that equivalently represents optical component vibrations as source vibrations. Their approach incorporated Bessel functions to quantify coherence degradation effects. Houghton et al. [10] implemented ray-tracing-based simulations that effectively capture geometric-optical vibration effects, particularly for focal spot stability analysis. While these methods have significantly advanced our understanding of beamline vibrations and their impacts on beam quality, a comprehensive wave-optics simulations that fully account for diffraction effects throughout the beamline still required. Such simulations would enable direct computation of time-evolving wavefronts, providing complete characterization of vibration-induced effects along the beamline including: (1) beam position fluctuations, (2) intensity distribution modifications, and (3) coherence degradation. However, conventional wave-optics methods face prohibitive computational challenges [11, 12], particularly in vibration analyses requiring extensive time-domain sampling, making full beamline simulations impractical for most practical applications.

To overcome these limitations, we propose a vibration mode propagation (VMP) method that combines three key innovations: First, vibrating wavefronts are decomposed into vibration modes using singular value decomposition (SVD). Second, instead of the vibrating wavefronts, these vibration modes were propagated through optical components. Third, time-domain vibrations are reconstructed through vibration modes superposition, eliminating the need for direct time-step integration. For actual beamline configuration, this approach achieves 40-fold computational acceleration compared to conventional wave-optics methods.

## 2. Model description

### 2.1 Vibration mode decomposition

We represent a vibrating wavefront as $E(x, y; t)$. For a time series from $t_0$ to $t_m$, the vibration operator $Vib$ generates a series of wavefronts:

$$Vib[E(x,y)] = [E(x,y;t_0), E(x,y;t_1), \dots, E(x,y;t_m)], \tag{1}$$

Using singular value decomposition (SVD) [13], the vibrating wavefront series can be decomposed as:

$$Vib(E) = U\Sigma V^+, \tag{2}$$

where $U$ is the matrix of vibration modes, $\Sigma$ is the singular value matrix, and $V$ is the evolution matrix of vibration parameters. Consequently, the wavefront at time $t_i$ can be

expressed as:

$$E(x, y; t_i) = \sum_n v_n(t_i) \times \rho_n \times \phi_n(x, y), \qquad (3)$$

where $\phi_n(x, y)$ is the $n^{\text{th}}$ vibration mode, $\rho_n$ is the corresponding singular value, and $v_n(t_i)$ is the vibration parameter at time $t_i$.

By propagating vibration modes instead of full wavefronts, we achieve significant computational efficiency in wave-optics simulations of beamline vibrations. And the efficiency of this method depends on the truncation index of vibration modes. To evaluate this, we analyzed a wavefront emitted from a single electron in an undulator at 8 keV [14]. A sinusoidal vibration with an amplitude of 10% of the wavefront size (8.6 $\mu$m FWHM) was applied along the horizontal direction. Figure 1a-e illustrates the decomposed vibration modes, with corresponding vibration parameters (Figure 1f) and singular values (Figure 1g). The cumulative occupation ratio of the first five vibration modes reaches 98.93% (Figure 1h).

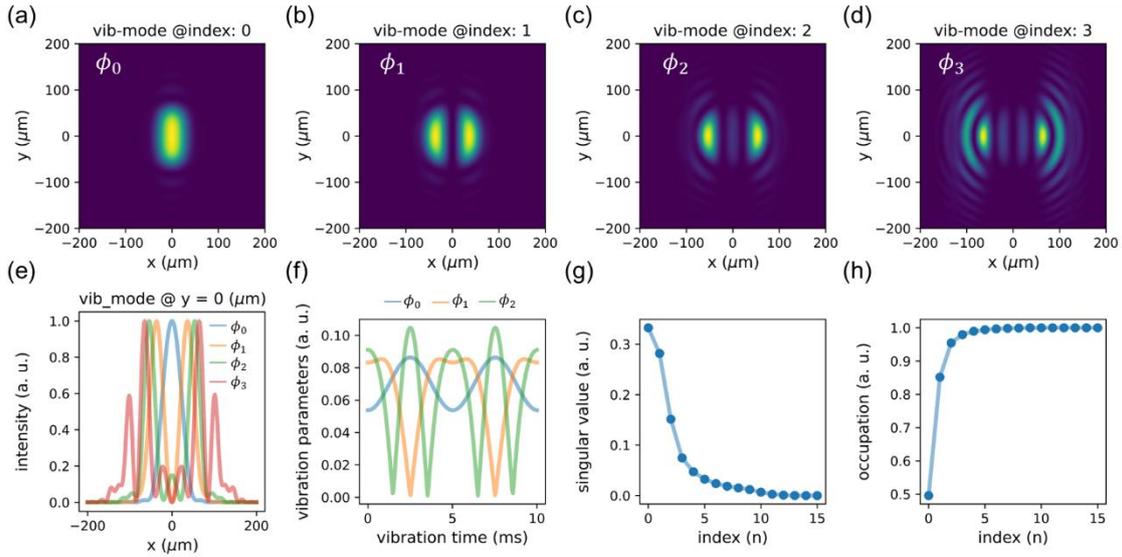

Figure 1. (a-e) The first four decomposed vibration modes $\phi_n(x, y)$ and their cross-sectional profiles at $y = 0$ $\mu$m.(e). The color blue, orange, green and red corresponding to the vibration modes with index = 0, 1, 2, and 3. (f) Time-dependent vibration parameters $v_n(t_i)$ for the first three vibration modes. The color blue, orange, and green corresponding to the vibration modes with index = 0, 1, and 2. (g) Singular values $\rho_n$ of the decomposed vibration modes. (h) Cumulative occupation ratio of the vibration modes

To quantify reconstruction accuracy based on Equation 3, we reconstructed the wavefront at its leftmost position using varying numbers of vibration modes. As shown in Figure 2, six vibration modes suffice to reconstruct the wavefront with a root-mean-square error (RMSE) of 0.4%, corresponding to a cumulative occupation ratio of 99.6%. These results demonstrate the high efficiency and accuracy of vibration mode propagation for wave-optics simulations.

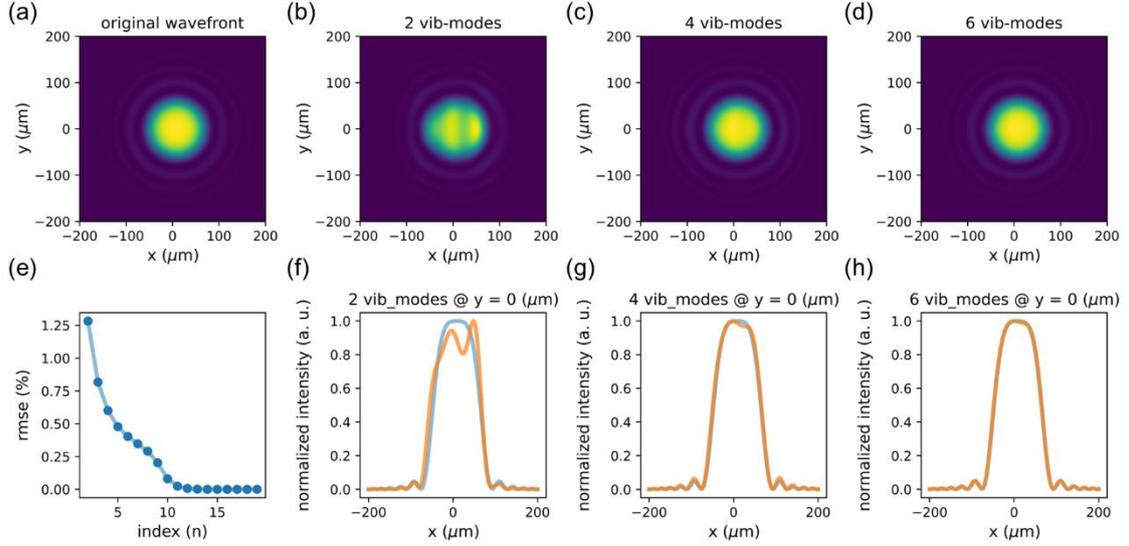

Figure 2. (a) The original wavefront. (b-d) The reconstructed wavefronts using 2, 4 and 6 vibration modes. (e) RMSE of the wavefront reconstruction using different number of vibration modes. (f-h) The cross-sectional profiles of the reconstructed wavefronts at $y = 0$ $\mu$m. The color blue and orange corresponding to the original and reconstructed wavefronts.

## 2.2 The vibration of optics

We categorize optical vibrations into two types: position vibrations and angular vibrations.

For position vibrations, we apply the Fourier shift theorem [15] to simulate wavefront displacements:

$$E(x + \Delta x, y + \Delta y) = F^{-1}[F[E(x,y)] \times e^{ik\Delta x} \times e^{ik\Delta y}], \qquad (4)$$

where $F$ and $F^{-1}$ denote the Fourier transform and inverse Fourier transform, $\Delta x$ and $\Delta y$ are the horizontal and vertical displacements, and $k$ is the X-ray wave vector.

For angular vibrations, the induced wavefront tilt $\theta_{\text{wfr}}$ creates an optical path difference:

$$E(x, y, \theta_{wfr}) = E(x, y, 0) \times e^{ik \times \sin(\theta_{wfr})\Delta d}, \tag{5}$$

where $\Delta d$ is the distance from point $(x,y)$ to the wavefront's vibration center. The relationship between $\theta_{wfr}$ and the optic's vibration angle $\theta_{optic}$ depends on the optic type. For refractive optics like compound refractive lenses (CRLs), $\theta_{wfr} = \theta_{optic}$. For reflective optics like Kirkpatrick-Baez (KB) mirrors, $\theta_{wfr} = 2\theta_{optic}$.

As shown in Figure 3a-b, subpixel and non-integral pixel shifts were accurately achieved using the Fourier shift theorem (pixel size 1.95 $\mu$m). To validate angular vibration simulations, we modeled wavefront focusing using CRL and KB mirrors with a 1:2 focus ratio. The mirrors were placed 20 m from the focus point, and a vibration angle of 1 $\mu$rad was applied. The resulting focus spot displacements were 20 $\mu$m for the CRL (Figure 3c) and 40 $\mu$m for the KB mirror (Figure 3d), consistent with theoretical predictions.

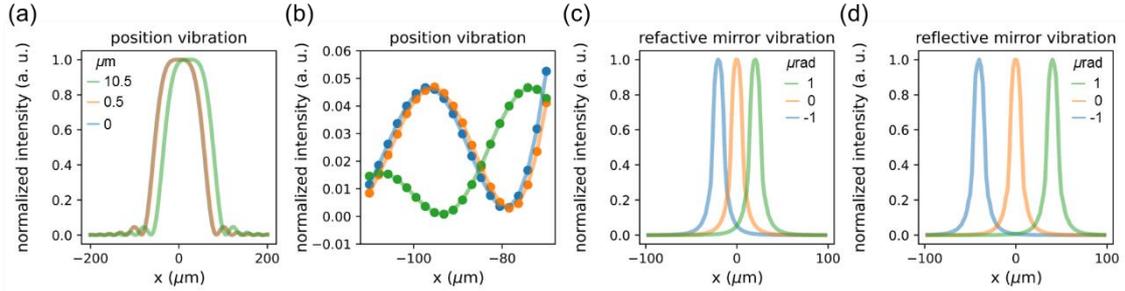

Figure 3. (a) The cross-sectional profiles of the vibrating wavefronts. (b) Detailed view of the vibrating wavefront profiles. (c) The vibrating focus using CRL. (d) The vibrating focus using KB mirror.

## 2.3. The wave-optics propagation of vibration

The vibration spectra of optical components in a beamline typically vary, complicating the propagation of wavefronts. Conventional approaches rely on extensive time-domain sampling to align these spectra, but this becomes complex and inefficient when different frequency vibrations are present. To address this, we developed a vibration mode-based propagation strategy that efficiently handles optics with different vibration spectra. Here, we describe the propagation process between two vibrating optics, $OP_1$ and $OP_2$, which can be extended to the entire beamline.

First, vibration modes at $OP_1$ are decomposed as detailed in Section 2.1. These modes are then propagated to $OP_2$ using standard wave-optics propagators, including the Fresnel diffraction integral (implemented via a double Fourier transform algorithm [16]), theangular spectrum method (ASM) [17], and the chirp-Z transform (CZT) [18]. The propagated wavefronts are expressed as:

$$E_{O_2}(x,y;t_{OP_1,i}) = P[E_{O_1}(x,y;t_{OP_1,i})] = P[\sum_n v_{OP_1,n}(t_{OP_1,i})\rho_{OP_1,n} \times \phi_{OP_1,n}(x,y)] =$$

$$\sum_n v_{OP_1,n}(t_{OP_1,i})\rho_{OP_1,n} \times P[\phi_{OP_1,n}(x,y)] = \sum_n v_{OP_1,n}(t_{OP_1,i})\rho_{OP_1,n} \times \phi_{OP_2,n}(x,y), \quad (6)$$

where $P$ denotes the propagator, and $\phi_{OP_2,n}(x,y)$ represents the propagated vibration modes on $OP_2$.

Next, the vibrations of $OP_2$ are applied to the propagated wavefronts:

$$Vib[E_{OP_2}(x,y;t_{OP_1,i})] = \sum_n v_{OP_1,n}(t_{OP_1,i})\rho_{OP_1,n} \times Vib[\phi_{OP_2,n}(x,y)], \quad (7)$$

Using Equation 3, $Vib[\phi_{OP_2,n}(x,y)]$ can be further decomposed into new vibration modes $\psi_{OP_2,n,m}(x,y)$:

$$Vib[\varphi_{OP_2,n}(x,y;t_{OP_2,j})] = \sum_m v_{OP_2,m}(t_{OP_2,j})\rho_{OP_2,m} \times \psi_{OP_2,n,m}(x,y), \quad (8)$$

Thus, the vibrating wavefronts after $OP_2$ at time $\tau$ are given by:

$$E_{OP_2}(x,y;\tau) = \sum_{n,m} v_{OP_1,n}(t_{OP_1,i})v_{OP_2,m}(t_{OP_2,j}) \times \rho_{OP_1,n}\rho_{OP_2,m} \times \psi_{OP_2,n,m}(x,y), \quad (9)$$

Since the vibration spectra of $OP_1$ and $OP_2$ differ, their time series $t_{OP1,i}$ and $t_{OP2,j}$ must be aligned to a common time series $\tau$. As only the vibration parameters $v$ is time-dependent, we interpolate them to $\tau$:

$$E_{OP_2}(x,y;\tau) = \sum_{n,m} v_{OP_1,n}^{interp}(\tau)v_{OP_2,m}^{interp}(\tau) \times \rho_{OP_1,n}\rho_{OPP_2,m} \times \psi_{OP_2,n,m}(x,y), \quad (10)$$

where $v_{OP_1,n}^{interp}(\tau)$ and $v_{OP_2,n}^{interp}(\tau)$ are the interpolated vibration parameters of the $n^{th}$ and $m^{th}$ modes at $OP_1$ and $OP_2$, respectively, aligned to the common time series $\tau$ using interpolation..

## 2.4. The model verification

To validate the developed VMP method, we compared its results with two benchmarks: (1) the conventional wavefront propagation method and (2) the theoretical decoherence model proposed by Shi et al [7]. A 1:2 focusing model was employed for this verification.

The wavefront emitted by a single electron served as the source, with an ideal focusing mirror placed 15 m downstream. The mirror had a focal length of 10 m, resulting in a focal plane 30 m from the mirror. Vibrations were introduced at both the source and the mirror, characterized by the following parameters: Source vibration amplitude and frequency are 5 $\mu$m and 100 Hz (horizontal direction); Mirror vibration amplitude and frequency are 5 $\mu$m and 10 Hz (vertical direction). Both vibrations followed a sinusoidal spectrum.

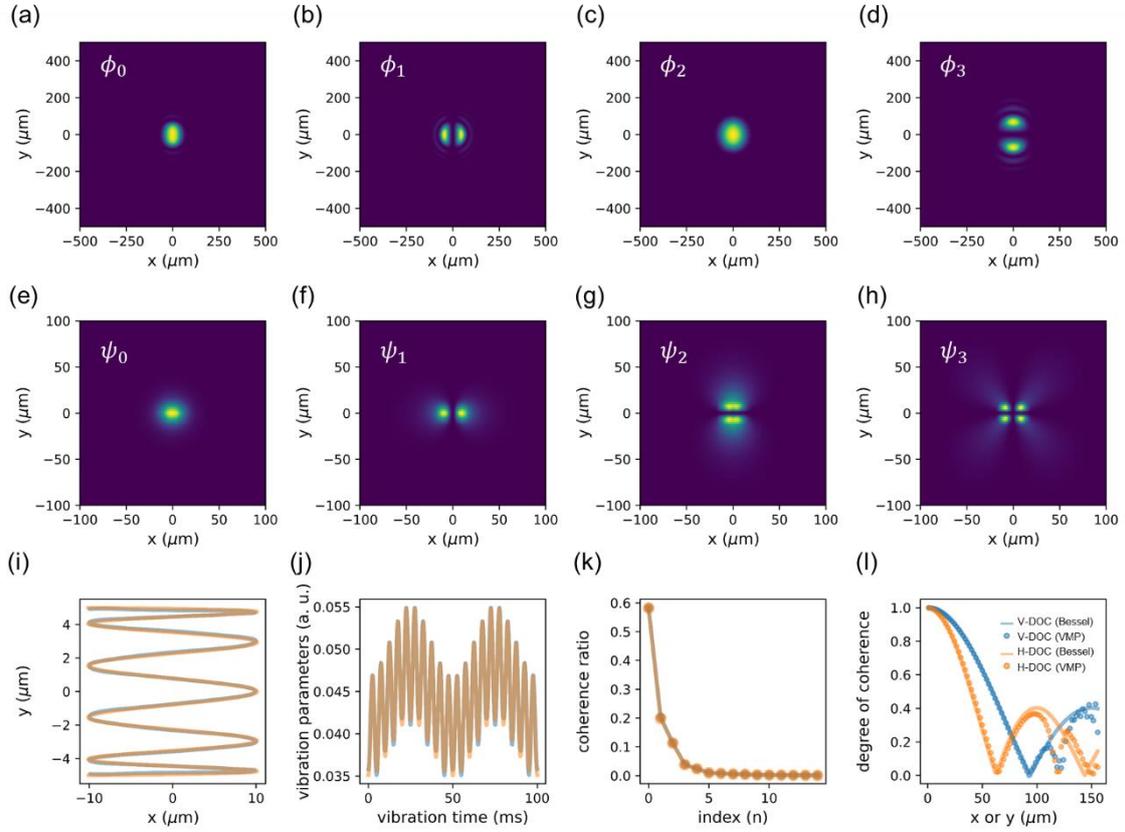

Figure 4. (a-d) The vibration modes of SR source. (e-h) The vibration modes at mirror. (i-k) The position, vibration parameters and coherent ratio of the vibrating focus. The color blue and orange corresponding to the results of VMP and conventional wavefront propagation methods. (l) The one-dimensional DOC along vertical (V-DOC) and horizontal (H-DOC) directions.

First, we propagated the wavefront through the vibration model using the conventional method. To align the time domains of the source and mirror, we set the simulation period to 0.1 seconds (for 10 Hz) with a sampling interval of 0.0002 seconds (50 points per 100 Hz cycle). The wavefront was propagated from the source to the focal plane at each time point, requiring 487 seconds of computation time.

Next, we applied the VMP method. The vibration modes $\phi_{source}(x,y)$ and their singular values were computed (Figure 4a-d). These modes were propagated to the mirror, where the mirror's vibrations were applied. The resulting vibrating modes $Vib[\phi_{mirror}(x,y)]$ were further decomposed into new modes $\psi_{mirror}(x,y)$ (Figure 4e-h). Finally, $\psi_{mirror}(x,y)$ were propagated to the focal plane, and the wavefronts were reconstructed, requiring only 56 seconds of computation time.

The positions of the focus spots, derived from Gaussian fitting, are compared in Figure 4i. The vibration parameters of the first vibration mode were analyzed (Figure 4j), demonstrating that the VMP method achieves the same accuracy as the conventional method. To evaluate decoherence effects, we compared the coherence ratios from both methods (Figure 4k). The results are consistent, confirming the reliability of the VMP method. Additionally, we validated the decoherence effects against the theoretical model of Shi et al [7]., which describes vibration-induced decoherence using the zero-order Bessel function:

$$|\mu(x)| = |J_0(2\pi A/\lambda D)|, \tag{11}$$

where $A$ is the vibration amplitude, $\lambda$ is the wavelength, $\mu(x)$ is the degree of coherence (DOC), and $D$ is the propagation distance. As shown in Figure 4l, the simulated DOC agrees with the theoretical predictions.

In summary, the VMP method effectively simulates beamline vibrations with significantly improved efficiency. In the 1:2 focusing model, the computation time was reduced by a factor of 10. As detailed below, the efficiency will be further increased for complex beamline configurations. This efficiency gain is attributed to two key factors: 1. the number of propagated vibration modes is much smaller than that of full wavefronts; 2. vibration combinations are achieved through one-dimensional evolution parameters, avoiding extensive time-domain sampling. These advantages make the VMP method a powerful tool for studying beamline vibrations.

# Simulations

## 3.1. The optical layout and vibration model of HXCS beamline

We applied the developed VMP method to analyze vibrations in the Hard X-ray Coherent Scattering (HXCS) beamline [14] at the High Energy Photon Source (HEPS). As a coherent beamline in HEPS Phase I, HXCS is designed to fully exploit the high brilliance and coherence of fourth-generation synchrotron radiation, supporting advanced techniques such as CDI, ptychography, and XPCS.

The optical layout of the HXCS beamline is shown in Figure 5a. Key optical components include: 1. SR source, the wavefront is calculated based on electron beam and undulator parameters; 2. Horizontal Double Crystal Monochromator (HDCM), positioned 40 meters from the source, the silicon (111) HDCM is oriented horizontally to enhance mechanical stability; 3. KB Mirrors, vertical KB (VKB) mirror at 79.62 meters and horizontal KB (HKB) mirror at 79.84 meters; 4. Coherent slit: placed upstream of the KB mirrors, the slit aperture is set to 160 × 398 $\mu m^2$ (horizontal × vertical) to adjust beam coherence; 5. Sample Position, Located 80 meters from the source, with the KB mirrors focusing the beam to approximately 100 nm for ptychography experiments.

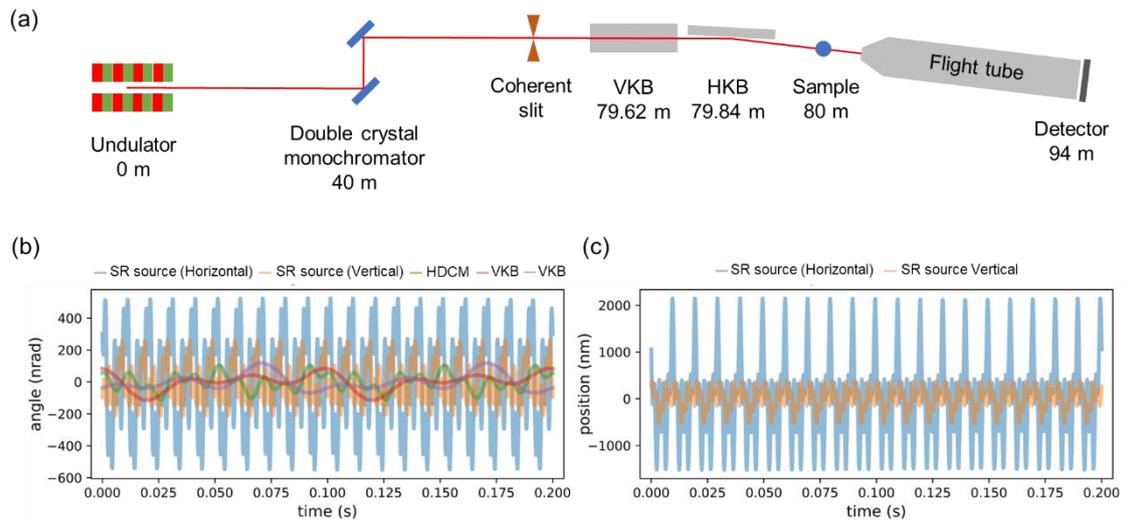

Figure 5. (a) The optical layout of the HXCS beamline. The angular (b) and position (c) vibration of the components.

The vibration model considers three primary vibration sources: the synchrotron source, HDCM, and KB mirrors. As detailed in Figure 5b-c, the positional and angular vibrations of these components were suggested by the HEPS optical mechanical group. The amplitude of the source vibrations was set to 10% of the source size, with frequencies approximately one order of magnitude higher than those of the HDCM and KB mirrors. HDCM and KB mirrors have lower-frequency vibrations, as specified in Figure 5b-c.

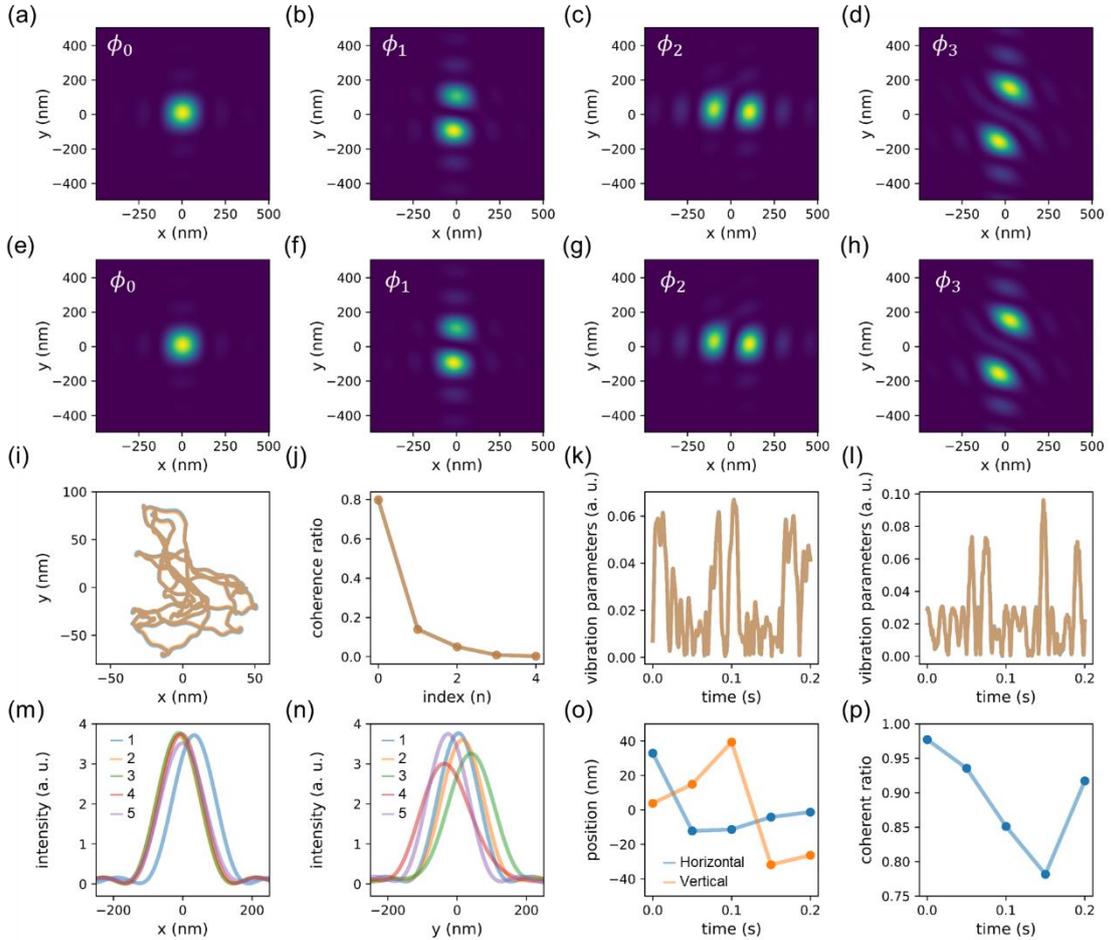

Figure 6. (a-d) Vibration modes $\phi_n$ simulated using the VMP method. (e-h) Vibration modes $\phi_n$ simulated using the conventional wavefront propagation method. (i) Comparison of focus positions obtained from both methods. (j) Coherence ratio as a function of vibration mode index. (k-l) Vibration parameters $v_n(t)$ for the second ($\phi_2$) and third ($\phi_3$) vibration modes. (m-n) Integrated intensity profiles along the horizontal (m) and vertical (n) directions, (o) focus positions over time, and (p) coherent ratio over time for the 50 Hz collection frequency.

To benchmarked the application of VMP method, wavefront propagation method was also performed for comparison. Wavefront propagation method required a time period of 0.2 seconds with a sampling interval of 0.2 milliseconds to capture the high-frequency source vibrations. VMP method used component-specific time periods and sampling intervals, significantly reducing computational complexity. The construction of the wave-optic vibration model was performed using the Coherent Analysis Toolbox (CAT) [14], with the first coherent mode of the light source (8 keV) serving as the SR source for the wave-optics simulation.

### 3.2. Vibration simulation of beamline

We analyzed the simulated results of the focus using the VMP method and compared them with those obtained from the conventional wavefront propagation method, as illustrated in Figure 6. The vibration modes (a-h), focus position (i), coherent ratio (j), and vibration parameters (k-l) from both methods were found to be consistent, validating the accuracy of the VMP method. The wavefront propagation method required 11.4 hours of computation time. Even for a 4$^{th}$ generation synchrotron source, where coherence mode decomposition methods could reduce computational complexity [12, 14], the resource requirements would still be enormous for wave-optics vibration simulation. In contrast, the VMP method completed the simulation in just 18 minutes, demonstrating a significant improvement in computational efficiency.

### 3.3. The effect of the vibration

We further analyzed the impact of beamline vibrations on coherent scattering experiments. In a typical coherent scattering process, a focused coherent X-ray probe $P(x,y)$ illuminates an object $O(x,y)$, producing an exit wavefront $P(x,y) \times O(x,y)$ [19]. The coherent diffraction intensity collected by the detector is given by:

$$I(x,y) = |F[P(x,y) \times O(x,y)]|^2, \qquad (12)$$

where $F$ denotes the Fourier transform, representing Fraunhofer diffraction from sample to detector. For an exposure time $\Delta t = [t_0, t_1, t_2, \ldots, t_n]$, the total diffraction intensity is:

$$I(x,y) = \sum_i I(x,y;t_i) = \sum_i |F[P(x,y;t_i) \times O(x,y)]|^2. \qquad (13)$$

Considering Equation 2, the vibrating X-ray probe can be expressed as:

$$P(x,y;t_i) = \sum_n v_n(t_i)\rho_n \times \varphi_n(x,y). \qquad (14)$$

Substituting this into the intensity equation yields:

$$I(x,y) = \sum_i \sum_n |v_n(t_i)\rho_n|^2 \times |F[\varphi_n(x,y) \times O(x,y)]|^2$$
$$= \sum_i \sum_n |v_n(t_i)|^2 \times |\rho_n|^2 \times |F[\varphi_n(x,y) \times O(x,y)]|^2. \quad (15)$$

Since $\sum_n |v_n(t_i)|^2 = 1$, the final intensity simplifies to:

$$I(x,y) = \sum_n |\rho_n|^2 \times |F[\varphi_n(x,y) \times O(x,y)]|^2. \quad (16)$$

This result demonstrates that vibrations within the detector's exposure time cause decoherence, while vibrations outside this period result in positional shifts. Thus, the detector's collection frequency critically determines the impact of beamline vibrations.

Therefore, we evaluated the HXCS beamline for three detector collection frequencies: 5 Hz, 50 Hz, and 500 Hz. 1. 5 Hz Collection Frequency: The beamline vibrations are averaged over the exposure time, reducing coherence by 21.4% (coherence ratio: 78.6%). The resulting coherence modes are the decomposed vibration modes (Figure 6a-d). 2. 50 Hz Collection Frequency: Both decoherence and positional effects are observed. As shown in Figure 6m and n, five groups of 0.2 s exposure results reveal variations in intensity distribution, focus spot positions (figure 6o), and coherent ratio (figure 6p) over time. 3. 500 Hz Collection Frequency: The beamline vibrations primarily cause positional shifts of the X-ray probe, with a root mean square error (RMSE) focus vibration of 19.4 nm.

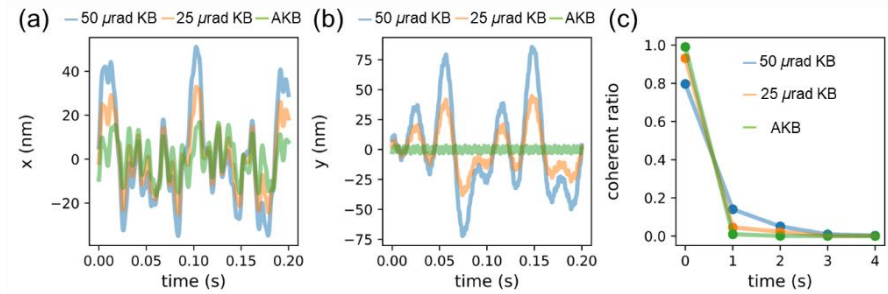

Figure 7. (a) Horizontal and (b) vertical positions of the vibrating focus spot over time. (c) Decoherence effect of the vibrating focus.

To mitigate vibration effects, we leveraged the high-efficiency VMP method to simulate and optimize the beamline design. For ptychography experiments, algorithms have been developed to handle partial coherence and positional shifts. First, we require the angular vibration of KB mirror to be 25 μrad, and the decoherence effect was reduced by 6.9%, and the position RMSE is 10.2 nm. Moreover, by replacing the traditional Kirkpatrick-Baez (KB) mirrors with advanced KB (AKB) mirrors [20], we significantly suppressed mirror vibrations, leaving only the source and double crystal monochromator (DCM) vibrations to consider. As shown in Figure 7, this optimization reduced the decoherence effect by 0.1% and decreased the positional RMSE to 4.4 nm. These results demonstrate that the stability of coherent beamlines can be greatly enhanced by adopting AKB mirrors.

**Conclusion**

In this study, we developed a highly efficient wave-optics vibration simulation method for $4^{th}$ generation synchrotron beamlines. By propagating vibration modes instead of full vibrating wavefronts, this method significantly reduces computational time and resource requirements. The accuracy of the method was validated through comparisons with theoretical models and conventional wavefront propagation methods, demonstrating excellent agreement. The applications of this method were demonstrated through practical examples. As shown in this study, the simulation time was reduced by more than 40 times compared to traditional methods, making it a powerful tool for beamline design and optimization. All the codes used in this study will be open source. Beyond synchrotron radiation sources, this method holds significant potential for application in other optical systems requiring wave-optics vibration analysis, such as optical inspection and lithography. The ability to efficiently simulate and analyze vibrations in high-precision optical systems opens new avenues for research and development in these fields.

## Acknowledgments.

This work was supported by High Energy Photon Source (HEPS), a major national science and technology infrastructure in China.

## Disclosures.

The authors declare no conflicts of interests.

## Data availability.

Data underlying the results presented in this paper are not publicly available at this time but may be obtained from the authors upon reasonable request.